\documentclass[twocolumn,showpacs]{revtex4}

\usepackage{graphicx}%
\usepackage{dcolumn}
\usepackage{amsmath}

\makeatletter
\def\btt#1{\texttt{\@backslashchar#1}}%
\DeclareRobustCommand\bblash{\btt{\@backslashchar}}%
\makeatother

\begin{document}

\preprint{HEP/123-qed}

\title[Short title]{Magnon-mediated NMR quantum gates in a 1-D antiferromagnet
}

\author{Atsushi Goto} 
 \email{GOTO.Atsushi@nims.go.jp}
\author{Tadashi Shimizu}%
\author{Kenjiro Hashi}%

\affiliation{Nanomaterials Laboratory, National Institute for Materials Science, Sakura, Tsukuba, 305-0003 Japan}

\date{\today}

\begin{abstract}
We propose a method of controlling a quantum logic gate in a solid-state NMR quantum computer.
A switchable inter-qubit coupling can be generated by using the longitudinal 
component of the Suhl-Nakamura interaction induced by a local singlet-triplet excitation 
in a 1-D antiferromagnet with a spin gap. 
\end{abstract}


\maketitle


\label{sec:introduction}
An implementation of a quantum computer (QC) has been a chief research target
in many fields of the materials science.
The concept of the QC is quite general, 
which allows us to suppose many quantum systems to be potential candidates for the QC.
Among them, a nuclear spin system is regarded as one of the most promising candidates
because of a good isolation from the environment and an excellent controllability 
by the well-established technique of the nuclear magnetic resonance (NMR).
Actually, the first 2-qubit QC's were implemented by solution NMR
\cite{chuang98, jones98},
which proved a great promise of the NMR-QC.
It is unfortunate, however, that the solution QC has a difficulty in its scalability,
because the number of available qubits
is limited by the number of nuclei in one molecule.
In order to increase the number of qubits systematically,
a solid-state (crystal) NMR-QC using a magnetic field gradient was proposed
\cite{yamaguchi99}.
Although there are some practical issues in the original proposal
\cite{hashi00}, 
its scalability is attractive to realize the multi-qubit QC.
It is our purpose to pursue the possibility of the solid-state NMR-QC 
by introducing new concepts into it.

One of the crucial processes in the QC implementation is to provide a controlled-NOT (CN) gate.
The CN gate is a two-qubit operation in which
a target qubit changes its logic according to the state of the control qubit.
It holds the essential part of the quantum computation,
where a quantum entanglement plays a crucial role.
In the NMR-QC, the entanglement is provided by an inter-nuclear coupling
so that it is the key issue to provide the appropriate inter-nuclear coupling.
In fact, the characteristics of the coupling considerably affect the fundamental properties of the QC, 
such as a design and a performance.
In particular, we find that a direct nuclear dipole coupling, 
which is regarded as a prospective inter-qubit coupling
\cite{yamaguchi99},
poses serious problems on the QC properties, as shown below.

From the practical point of view,
the following two properties are the crucial requisites for the inter-qubit coupling.
One is to be capable of switching. 
The coupling needs to be strong enough to finish the logic gate in an appropriate time scale,
but to be removable whenever unnecessary.
The nuclear dipole coupling is fortunate 
in that it is available whenever nuclear spins are put close to each other,
while unfortunate in that the decouplings 
for a huge number of inter-nuclear couplings are formidable.
It is desirable that the coupling be switched-on only when necessary.
The other property is to reach rather long range
so that qubits in the field gradient can be distinguished 
from each other in the frequency domain.  
The larger inter-qubit distance in the real space makes 
the frequency difference between adjacent NMR lines wider in the frequency domain, so that
the constraint on the field gradient can be relieved by using a long-range coupling.
Unfortunately, the nuclear dipole coupling reaches at most a few lattice points
\cite{goto01},
which motivates us to invoke some long-range indirect interactions via electrons.

In this letter, we present a switchable inter-qubit coupling 
realized by the Suhl-Nakamura (SN) interaction mediated by magnons
\cite{suhl59, nakamura58}
in a 1-D antiferromagnet with a spin gap.
The coupling is switched on selectively by local magnon excitations across the spin gap,
which works as a switch for the CN gate.
The proposed system is intuitive
in that the computation starts in a silent environment rather than the turbulence of interactions,
which simplifies the designs of the logic gates.
Moreover, since the SN interaction reaches rather long distance, 
one can take the inter-qubit distance longer 
and relieve the constraint on the field gradient significantly.

\begin{figure}
\includegraphics[scale=0.6]{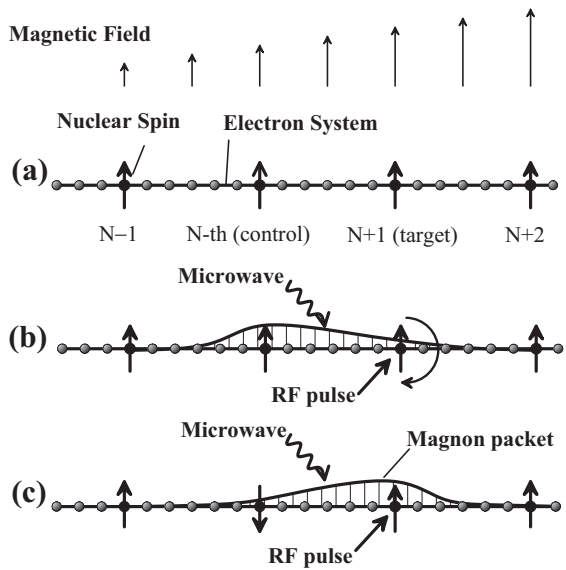}
\caption{
Schematic illustration of the CN-gate switching
in a 1D antiferromagnet in a magnetic field gradient.
The $I=0$ and $1/2$ (= qubits) nuclei are shown by circles and arrows, respectively.
(a) Before the microwave irradiation, all the inter-nuclear couplings are switched off.
(b), (c) A magnon packet (hatched part) is excited between the control 
and the target qubits, which creates the entanglement between them.
The additional magnetic field produced by the packet at the target qubit 
depends on the state of the control qubit ($h=h_{\rm tr} \pm h_{\rm SN}$)
corresponding to the NMR frequencies of $\omega_{\pm}=\gamma_n(H_0+h_{\rm tr} \pm h_{\rm SN})$.
The rf field with the frequency $\omega_{-}$ can rotate the target spin
in the case of (b), but not in (c).
The states of (b) and (c) are superposed in the actual computation.
}
\label{j-switch}
\end{figure}

\label{sec:model}
The main idea of the gate switching is illustrated in Fig.~\ref{j-switch}.
Suppose a device with an antiferromagnetic electron spin chain 
placed in a magnetic field gradient produced by a micromagonet 
fabilicated outside the spin chain. 
The spin chain is supposed to have a singlet ground state 
($\mid s s_z\rangle$=$\mid0 0\rangle$)
with a finite gap to the lowest triplet branch of the magnon modes 
because of a quantum many body effect.
Examples can be found in spin ladder and Haldane systems.
One could also use dimer and spin-Peierls systems.
Suppose that nuclear spins ($I=1/2$) serving as qubits can be placed periodically, 
e.g., every 10$a$ ($a$: lattice spacing),
with the rest of the sites be occupied with $I=0$ nuclei.
The field gradient enables us to access each qubit individually 
by tuning the NMR frequency.
The rather long distance between qubits is effective 
both to diminish the nuclear dipole couplings between qubits
and to distinguish each qubit in the frequency domain.

Since there are no unpaired electron spins in the ground state,
the interactions between qubits are absent at low enough $T$,
which means that each qubit is isolated from other qubits 
as well as the environment (= electron spin system).
Then, the inter-qubit coupling needed for the CN-gate is switched on 
through the SN interaction by the $k=0$ triplet magnons, 
which are created locally by the singlet-triplet excitation across the spin gap.
Although the excitation is primarily forbidden for the usual electric dipolar transition, 
it often becomes possible in the actual systems because of some higher order effects.
\cite{brill94,lu91}

\begin{figure}[t]
\includegraphics[scale=0.3]{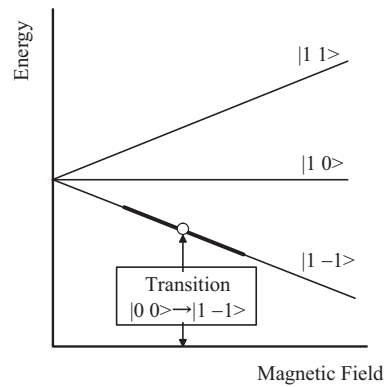}
\caption{Energy diagrams for the three triplet branches 
($\mid 1 1 \rangle, \mid 1 0 \rangle$ and $\mid 1 -1 \rangle$) 
of $k = 0$ magnon modes 
against the singlet ($\mid 0 0 \rangle$) as a function of magnetic field.
In a linear field gradient, the horizontal axis also corresponds to the position along the chain.
The thick line on the $\mid1-1\rangle$ branch shows the region where the actual chain is located, 
and an open circle is the spot where the transition occurs.}
\label{esr}
\end{figure}

The position of the excited magnons along the chain can be specified in the field gradient.
The energy diagrams of the $k=0$ magnon excitations 
in the magnetic field are shown in Fig. ~\ref{esr} 
\cite{tachiki70, brill94}.
The magnon excitation energy to $\mid1-1\rangle$ are different position by position 
along the chain because of the field gradient,
which provides us with a spatial resolution of the magnon excitation region.
The microwave frequency is adjusted
so as to specify the pair of qubits to be coupled.
The process creates a packet of the superposition of 
$\mid 0 0\rangle$ and $\mid1 -1\rangle$
\cite{brill94},
corresponding to a soliton-like magnon excitation with $k \sim 0$.

The packet is localized on the chain due, once again, to the magnetic field gradient.
A mismatch in the magnon excitation energies between adjacent regions along the chain 
prohibits the packet from moving to the lower field region. 
On the other hand, 
the continuum excitations near the one-magnon excitations at $k$=0 is absent,
so that it is difficult for the packet to move to the higher field region
\cite{augier98, uhrig96, garrett97, yu00, grenier00}
unless the process of the energy release by phonon emissions is considerable.
Consequently, the magnons are confined in the region where they are excited, 
and the SN interaction is produced only between the qubit pair of interest. 

The triplet states make an additional field ($h_{\rm tr}$) at the target qubit,
which should be distinguished from that from the control qubit 
via the SN interaction ($h_{\rm SN}$).
$h_{\rm tr}$ can be measured 
by observing the NMR frequency shift of the target qubit during the microwave irradiation, 
while saturating the control qubit by applying the rf field continuously.
The saturation of the control qubit results in $h_{\rm SN}=0$ at the target qubit site, 
so that the observed {\it shift} directly corresponds to $h_{\rm tr}$.
 
The SN coupling between the two qubits $H_{\rm SN}$ is given by $H_{\rm SN}=W_{ij}I_i^zI_j^z$ with,
\begin{equation}
 W_{ij} = \left (\frac{\gamma_nA}{N} \right )^2
\sum_{k \neq k^{'}} \frac{n_k-n_{k^{'}}}{\epsilon_{k^{'}}-\epsilon_k} \cos \{(k-k^{'}) r_{ij}\}.
\label{longitudinal-SN}
\end{equation}
Here, $n_k$ and $\epsilon_k$ are, respectively, the population and the energy 
of the magnon with the wave number $k$,
and $r_{ij}$ is the distance between the two qubits of interest.
\cite{zevin75}
$A$ and $N$ are the hyperfine coupling constant
and the number of sites (including the $I=0$ sites) inside the packet, respectively.

Suppose the magnon dispersion of the spin ladder, as an example, in the form,
$\epsilon(k_n) = C+J(j_1-j_1^3/4) \cos(k_n)+ ...$,
where $k_n  = n\pi /N$, $j_1 = J_1/J$ 
with $J_1$ and $J$ being the intra- and inter-chain exchange interactions, 
and $C$ is the part independent of $k_n$
\cite{muller00}.
Recalling that the microwave irradiation excites only the $k=0$ magnons, 
i.e., $n(k)=0$ for $k \neq 0$, 
the range function $W_{ij}$ can be calculated as,
\begin{eqnarray}
 W_{ij} & = & \left (\frac{\gamma_n A_{\parallel}}{N} \right )^2
\sum_{n=1}^{N} \frac{2 n(0)}{\epsilon(k_n)-\epsilon(0)} \cos (k_n r_{ij}) \nonumber\\
 & = & \frac{2 \gamma_n^2 A_{\parallel}^2 \{n(0)/N\} }{J(j_1-\frac{1}{4}j_1^3)N}
 \sum_{n=1}^N \frac{\cos (k_n r_{ij})}{\cos (k_n)-1}.
\end{eqnarray}
Assuming $N$ = 20, $r_{ij}$ = 10$a$, 
$A_{\parallel}$ = 100 kOe/$\mu_B$, 
$\gamma_n/(2 \pi)$ = 4.3 MHz/kOe ($^{1}$H as an example), $J$ = 50 K, $j_2$ = 0.2 
and $n(0)/N = 0.01$,
one obtains $W_{ij}$ = 15 kHz,
which is the same order of magnitude as
the nuclear dipole coupling acting between nuclei 3 \AA \hspace{1mm} apart from each other
\cite{yamaguchi99, ladd00}, 
and one to three orders of magnitude greater than the J-couplings used in the solution NMR-QC's
\cite{chuang98, jones98}.
Note that $W_{ij}$ can be controlled by the microwave intensity via $n(0)$,
i.e, $n(0)$ is determined by the balance between excitation and relaxation;
${d n(0)}/{dt}=W_{\rm ex}-{n(0)}/{T_s},$
with $W_{\rm ex}$ being the transition probability per unit time 
and $T_s$ the magnon lifetime.
In the steady state, $d n(0)/dt=0$, so that $n(0) = W_{\rm ex} T_s$.

\begin{figure}[t]
\includegraphics[scale=0.6]{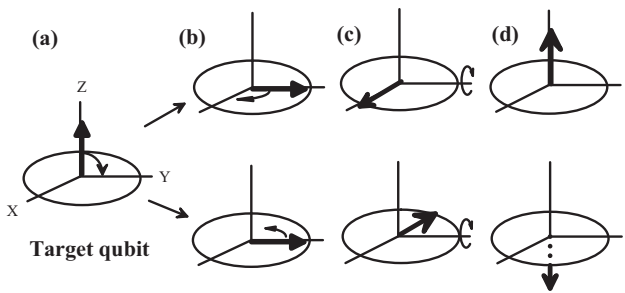}
\caption{
Schematic illustration of the CN gate 
in the rotating frame of the target qubit. 
Upper and lower figures correspond to the up (0) and the down (1)
states of the control qubit. 
(a) In the beginning, the target qubit is pointing to the Z-direction.
The spin is rotated by a $\pi/2$ pulse 
with the frequency of $\omega=\gamma_n(H_0+h_{tr})$ around the -X axis.
(b) The spin starts to turn in the XY plane 
due to the additional field caused by the spin of the control qubit. 
(c) The spin turns by $\pm 90^o$ to +X or -X direction in the XY plane. 
(d) The spin is rotated again by the $\pi/2$ pulse around the -Y axis.  
The target qubit is hereby
controlled according to the state of the control qubit.
}
\label{CN-gate}
\end{figure}
The CN gate is achieved by an NMR technique shown in Fig. \ref{CN-gate}.
After the gate operation, the couping is switched-off 
by shutting-off the microwave irradiation. 
The triplet state relaxes to the singlet ground state by the time scale of $T_s$.
Through these processes, the microwave irradiation works as a switch for the CN gate.

In summary, we have described the magnon-mediated quantum gate
in a 1-D antiferromagnet.
The switching capability of the gate is useful 
in designing the logic gate of the solid-state NMR-QC, 
whereas the long-range nature of the coupling significantly relieves 
the constraint on the field gradient required to distinguish each qubits.
We are indebted to M. Kitagawa for fruitful discussion. 
This work has been supported by CREST of JST (Japan Science and Technology Corporation).

\bibliography{J-letter}

\begin{thebibliography}{18}
\expandafter\ifx\csname natexlab\endcsname\relax\def\natexlab#1{#1}\fi
\expandafter\ifx\csname bibnamefont\endcsname\relax
  \def\bibnamefont#1{#1}\fi
\expandafter\ifx\csname bibfnamefont\endcsname\relax
  \def\bibfnamefont#1{#1}\fi
\expandafter\ifx\csname citenamefont\endcsname\relax
  \def\citenamefont#1{#1}\fi
\expandafter\ifx\csname url\endcsname\relax
  \def\url#1{\texttt{#1}}\fi
\expandafter\ifx\csname urlprefix\endcsname\relax\def\urlprefix{URL }\fi
\providecommand{\bibinfo}[2]{#2}
\providecommand{\eprint}[2][]{\url{#2}}

\bibitem[{\citenamefont{Chuang et~al.}(1998)\citenamefont{Chuang, Gershenfeld,
  and Kubinec}}]{chuang98}
\bibinfo{author}{\bibfnamefont{I.~L.} \bibnamefont{Chuang}},
  \bibinfo{author}{\bibfnamefont{N.}~\bibnamefont{Gershenfeld}},
  \bibnamefont{and} \bibinfo{author}{\bibfnamefont{M.}~\bibnamefont{Kubinec}},
  \bibinfo{journal}{Phys.\ Rev.\ Lett.} \textbf{\bibinfo{volume}{80}},
  \bibinfo{pages}{3408} (\bibinfo{year}{1998}).

\bibitem[{\citenamefont{Jones and Mosca}(1998)}]{jones98}
\bibinfo{author}{\bibfnamefont{J.~A.} \bibnamefont{Jones}} \bibnamefont{and}
  \bibinfo{author}{\bibfnamefont{M.}~\bibnamefont{Mosca}}, \bibinfo{journal}{J.
  Chem.\ Phys..} \textbf{\bibinfo{volume}{109}}, \bibinfo{pages}{1648}
  (\bibinfo{year}{1998}).

\bibitem[{\citenamefont{Yamaguchi and Yamamoto}(1999)}]{yamaguchi99}
\bibinfo{author}{\bibfnamefont{F.}~\bibnamefont{Yamaguchi}} \bibnamefont{and}
  \bibinfo{author}{\bibfnamefont{Y.}~\bibnamefont{Yamamoto}},
  \bibinfo{journal}{Appl.\ Phys.\ A} \textbf{\bibinfo{volume}{68}},
  \bibinfo{pages}{1} (\bibinfo{year}{1999}).

\bibitem[{\citenamefont{Hashi et~al.}(2000)\citenamefont{Hashi, Shimizu, Goto,
  Kitazawa, Kido, and Suzuki}}]{hashi00}
\bibinfo{author}{\bibfnamefont{K.}~\bibnamefont{Hashi}},
  \bibinfo{author}{\bibfnamefont{T.}~\bibnamefont{Shimizu}},
  \bibinfo{author}{\bibfnamefont{A.}~\bibnamefont{Goto}},
  \bibinfo{author}{\bibfnamefont{H.}~\bibnamefont{Kitazawa}},
  \bibinfo{author}{\bibfnamefont{G.}~\bibnamefont{Kido}}, \bibnamefont{and}
  \bibinfo{author}{\bibfnamefont{T.}~\bibnamefont{Suzuki}},
  \bibinfo{journal}{Appl.\ Phys.\ A} \textbf{\bibinfo{volume}{70}},
  \bibinfo{pages}{359} (\bibinfo{year}{2000}).

\bibitem[{\citenamefont{Goto et~al.}(2002)\citenamefont{Goto, Shimizu, Miyabe,
  Hashi, Kitazawa, Kido, Shimamura, and Fukuda}}]{goto01}
\bibinfo{author}{\bibfnamefont{A.}~\bibnamefont{Goto}},
  \bibinfo{author}{\bibfnamefont{T.}~\bibnamefont{Shimizu}},
  \bibinfo{author}{\bibfnamefont{R.}~\bibnamefont{Miyabe}},
  \bibinfo{author}{\bibfnamefont{K.}~\bibnamefont{Hashi}},
  \bibinfo{author}{\bibfnamefont{H.}~\bibnamefont{Kitazawa}},
  \bibinfo{author}{\bibfnamefont{G.}~\bibnamefont{Kido}},
  \bibinfo{author}{\bibfnamefont{K.}~\bibnamefont{Shimamura}},
  \bibnamefont{and} \bibinfo{author}{\bibfnamefont{T.}~\bibnamefont{Fukuda}},
  \bibinfo{journal}{Appl.\ Phys.\ A} \textbf{\bibinfo{volume}{74}},
  \bibinfo{pages}{73} (\bibinfo{year}{2002}).

\bibitem[{\citenamefont{Suhl}(1959)}]{suhl59}
\bibinfo{author}{\bibfnamefont{H.}~\bibnamefont{Suhl}}, \bibinfo{journal}{J.\
  Phys.\ Radium} \textbf{\bibinfo{volume}{20}}, \bibinfo{pages}{333}
  (\bibinfo{year}{1959}).

\bibitem[{\citenamefont{Nakamura}(1958)}]{nakamura58}
\bibinfo{author}{\bibfnamefont{T.}~\bibnamefont{Nakamura}},
  \bibinfo{journal}{Prog.\ Theor.\ Phys.} \textbf{\bibinfo{volume}{20}},
  \bibinfo{pages}{542} (\bibinfo{year}{1958}).

\bibitem[{\citenamefont{Brill et~al.}(1994)\citenamefont{Brill, Boucher,
  Voiron, Dhalenne, Revcolevschi, and Renard}}]{brill94}
\bibinfo{author}{\bibfnamefont{T.~M.} \bibnamefont{Brill}},
  \bibinfo{author}{\bibfnamefont{J.~P.} \bibnamefont{Boucher}},
  \bibinfo{author}{\bibfnamefont{J.}~\bibnamefont{Voiron}},
  \bibinfo{author}{\bibfnamefont{G.}~\bibnamefont{Dhalenne}},
  \bibinfo{author}{\bibfnamefont{A.}~\bibnamefont{Revcolevschi}},
  \bibnamefont{and} \bibinfo{author}{\bibfnamefont{J.~P.}
  \bibnamefont{Renard}}, \bibinfo{journal}{Phys.\ Rev.\ Lett.}
  \textbf{\bibinfo{volume}{73}}, \bibinfo{pages}{1545} (\bibinfo{year}{1994}).

\bibitem[{\citenamefont{Lu et~al.}(1991)\citenamefont{Lu, Tuchendler, von
  Ortenberg, and Renard}}]{lu91}
\bibinfo{author}{\bibfnamefont{W.}~\bibnamefont{Lu}},
  \bibinfo{author}{\bibfnamefont{J.}~\bibnamefont{Tuchendler}},
  \bibinfo{author}{\bibfnamefont{M.}~\bibnamefont{von Ortenberg}},
  \bibnamefont{and} \bibinfo{author}{\bibfnamefont{J.~P.}
  \bibnamefont{Renard}}, \bibinfo{journal}{Phys.\ Rev.\ Lett.}
  \textbf{\bibinfo{volume}{67}}, \bibinfo{pages}{3716} (\bibinfo{year}{1991}).

\bibitem[{\citenamefont{Tachiki and Yamada}(1970)}]{tachiki70}
\bibinfo{author}{\bibfnamefont{M.}~\bibnamefont{Tachiki}} \bibnamefont{and}
  \bibinfo{author}{\bibfnamefont{T.}~\bibnamefont{Yamada}},
  \bibinfo{journal}{J. Phys.\ Soc.\ Jpn.} \textbf{\bibinfo{volume}{28}},
  \bibinfo{pages}{1413} (\bibinfo{year}{1970}).

\bibitem[{\citenamefont{Augier and Poilblanc}(1998)}]{augier98}
\bibinfo{author}{\bibfnamefont{D.}~\bibnamefont{Augier}} \bibnamefont{and}
  \bibinfo{author}{\bibfnamefont{D.}~\bibnamefont{Poilblanc}},
  \bibinfo{journal}{Eur.\ Phys.\ J.\ B} \textbf{\bibinfo{volume}{1}},
  \bibinfo{pages}{19} (\bibinfo{year}{1998}).

\bibitem[{\citenamefont{Uhrig and Schulz}(1996)}]{uhrig96}
\bibinfo{author}{\bibfnamefont{G.~S.} \bibnamefont{Uhrig}} \bibnamefont{and}
  \bibinfo{author}{\bibfnamefont{H.~J.} \bibnamefont{Schulz}},
  \bibinfo{journal}{Phys.\ Rev.\ B} \textbf{\bibinfo{volume}{54}},
  \bibinfo{pages}{R9624} (\bibinfo{year}{1996}).

\bibitem[{\citenamefont{Garrett et~al.}(1997)\citenamefont{Garrett, Nagler,
  Tennant, Sales, and Barnes}}]{garrett97}
\bibinfo{author}{\bibfnamefont{A.~W.} \bibnamefont{Garrett}},
  \bibinfo{author}{\bibfnamefont{S.~E.} \bibnamefont{Nagler}},
  \bibinfo{author}{\bibfnamefont{D.~A.} \bibnamefont{Tennant}},
  \bibinfo{author}{\bibfnamefont{B.~C.} \bibnamefont{Sales}}, \bibnamefont{and}
  \bibinfo{author}{\bibfnamefont{T.}~\bibnamefont{Barnes}},
  \bibinfo{journal}{Phys.\ Rev.\ Lett.} \textbf{\bibinfo{volume}{79}},
  \bibinfo{pages}{745} (\bibinfo{year}{1997}).

\bibitem[{\citenamefont{Yu and Haas}(2000)}]{yu00}
\bibinfo{author}{\bibfnamefont{W.}~\bibnamefont{Yu}} \bibnamefont{and}
  \bibinfo{author}{\bibfnamefont{S.}~\bibnamefont{Haas}},
  \bibinfo{journal}{Phys.\ Rev.\ B} \textbf{\bibinfo{volume}{62}},
  \bibinfo{pages}{344} (\bibinfo{year}{2000}).

\bibitem[{\citenamefont{Grenier et~al.}(2000)\citenamefont{Grenier, Regnault,
  Lorenzo, Boucher, Hiess, Dhalenne, and Revcolevschi}}]{grenier00}
\bibinfo{author}{\bibfnamefont{B.}~\bibnamefont{Grenier}},
  \bibinfo{author}{\bibfnamefont{L.~P.} \bibnamefont{Regnault}},
  \bibinfo{author}{\bibfnamefont{J.~E.} \bibnamefont{Lorenzo}},
  \bibinfo{author}{\bibfnamefont{J.~P.} \bibnamefont{Boucher}},
  \bibinfo{author}{\bibfnamefont{A.}~\bibnamefont{Hiess}},
  \bibinfo{author}{\bibfnamefont{G.}~\bibnamefont{Dhalenne}}, \bibnamefont{and}
  \bibinfo{author}{\bibfnamefont{A.}~\bibnamefont{Revcolevschi}},
  \bibinfo{journal}{Phys.\ Rev.\ B} \textbf{\bibinfo{volume}{62}},
  \bibinfo{pages}{12206} (\bibinfo{year}{2000}).

\bibitem[{\citenamefont{Zevin and Kaplan}(1975)}]{zevin75}
\bibinfo{author}{\bibfnamefont{V.}~\bibnamefont{Zevin}} \bibnamefont{and}
  \bibinfo{author}{\bibfnamefont{N.}~\bibnamefont{Kaplan}},
  \bibinfo{journal}{Phys.\ Rev.\ B} \textbf{\bibinfo{volume}{12}},
  \bibinfo{pages}{4604} (\bibinfo{year}{1975}).

\bibitem[{\citenamefont{Muller and Mikeska}(2000)}]{muller00}
\bibinfo{author}{\bibfnamefont{M.}~\bibnamefont{Muller}} \bibnamefont{and}
  \bibinfo{author}{\bibfnamefont{H.-J.} \bibnamefont{Mikeska}},
  \bibinfo{journal}{J. Phys.\ Condens.\ Matter} \textbf{\bibinfo{volume}{12}},
  \bibinfo{pages}{7633} (\bibinfo{year}{2000}).

\bibitem[{\citenamefont{Ladd et~al.}(2000)\citenamefont{Ladd, Goldman,
  Yamaguchi, and Yamamoto}}]{ladd00}
\bibinfo{author}{\bibfnamefont{T.~D.} \bibnamefont{Ladd}},
  \bibinfo{author}{\bibfnamefont{J.~R.} \bibnamefont{Goldman}},
  \bibinfo{author}{\bibfnamefont{F.}~\bibnamefont{Yamaguchi}},
  \bibnamefont{and} \bibinfo{author}{\bibfnamefont{Y.}~\bibnamefont{Yamamoto}},
  \bibinfo{journal}{Appl.\ Phys.\ A} \textbf{\bibinfo{volume}{71}},
  \bibinfo{pages}{27} (\bibinfo{year}{2000}).

\end{thebibliography}

\end{document}